\documentclass[multi]{cambridge7A} 
\usepackage{times}
\usepackage{natbib}
\usepackage{rotating}
\usepackage{floatpag}
  \rotfloatpagestyle{empty} 
\usepackage{graphicx}
\usepackage{epstopdf}

\begin{document}

\author{Steven D. Kawaler\footnotemark}
\chapter{Learning Physics from the Stars: \\ Its All in the Coefficients}
\footnotetext[1]{Department of Physics and Astronomy, Iowa State University, Ames, IA USA 50011}

\section{Overview and basic discussion of equations of stellar structure}

This section is intended merely as a reminder of things you have already learned (or will soon learn) in a first course in stellar structure and evolution.  Here, we will introduce, and in some cases derive, the four basic equations of stellar structure.  If you'd like additional details, by all means consult one of the standard texts in this subject - I recommend without hesitation the text by \citet{HKT04}.

\subsection{Dependent and Independent Variables}

In solving the equations that describe the conditions within stars, we need to decide on a coordinate system.  A reasonable simplification of the equations arises if we assume that stars are spherically symmetric.  WIth that assumption, a single positional variable will suffice as the independent variable.  Our choice could be $r$, the distance from the center of the star, or $m_r$, the mass contained within a shell of radius $r$.  For some purposes we may use $r$ and for others, $m_r$.  Their values are related through the equation of mass conservation (or continuity):
\begin{equation}
\frac{dm_r}{dr} = 4 \pi r^2 \rho(r),
\label{continuity}
\end{equation}
where $\rho(r)$ is the mass density at position $r$.

As for the quantities that we would like to use to describe the physical conditions within the star at each position, we will define them as follows:
\begin{itemize}
 \item{velocity: $v$ (=0 for a hydrostatic star)}
 \item{density: $\rho$ or $n=N_A \rho / \mu$ }
 \item{pressure: $P$}
 \item{temperature: $T$}
 \item{chemical composition (mass fraction): $X_i$ }
 \item{ion / charge balance: $y_i$, $n_e$ } 
 \item{internal energy: $U$}
 \item{entropy: $S$}
 \item{heat flow parameters / cross sections: $\kappa_{\rm rad}, \kappa_{\rm cond}$ }
 \item{energy flow: $L_{\rm r}, F_{\rm conv}$ }
 \item{energy generation / loss: $\epsilon_{\rm nuc}, \epsilon_{\nu}$}
\end{itemize}

These quantities provide a fairly complete description of the state of the stellar interior, but they clearly are not independent of one another.  In fact, one can show that only four dependent variables at each point in a model suffice to fully describe the conditions there.  For computation of stellar structure and evolution, those quantities are usually chosen as $r$ (or $m_r$), $P$, $T$, and $L_r$.

Of course, the other quantities are important, but they can be derived or computed given the main four quantities.  That said, the computation of a stellar model requires knowledge of the values of all of the above quantities.  They appear in the {\em coefficients} of the equations, and contain all of the physics of relevance to stellar interiors.  Thus they provide the direct connection between fundamental physics and the appearance and behavior of the stars.

Their appearance in the coefficients of the equations of stellar structure allows us to further categorize what we need in terms of the primary {\em mechanical} quantities $r$ (or $m_r$) and $P$, and the primary {\em thermal} quantities $T$ and $L_r$.  Necessary extra information is given by the composition ($X_i$).  Various branches of physics provide the tools do determine important {\em derived} quantities include equation--of--state parameters $\rho$, $\mu$, $U$, and $S$, from which we can calculate $\nabla_{\rm ad}$, specific heats, and other thermodynamic quantities.  From atomic physics we can compute $y_i$ and $n_e$, and the opacities $\kappa_{\rm rad}$ and $\kappa_{\rm cond}$.  Nuclear physics provides the tools for computing $\epsilon_{\rm nuc}$ and $\epsilon_{\rm neut}$, and hydrodynamics gives us hope for computing $F_{\rm conv}$.

\subsection{The Equations of Stellar Structure}

Let's now see how the above set of parameters appears when we write down the equations of stellar structure.  This is possible to do in a compact form only if we expect that the assumption of spherical symmetry everywhere is a good one (so that all quantities are functions only of $r$), and that the star that we are modeling is in hydrostatic equilibrium (ensuring that time derivatives are zero).  Then and only then do the equations take on the form that is used, almost universally, used to produce stellar models for asteroseismological (and indeed most other) purposes.

\subsubsection{Mass Conservation}

The first equation has already been introduced above; equation \ref{continuity} ensures that the mass and radius are consistent with the density from the center to the surface.  Equation \ref{continuity} also serves as a coordinate transformation between using $m_r$ and $r$ as independent variables.  Consider a general quantity $Z$ (which could stand for temperature, pressure, or something else).  Then the differential equation describing the dependence of $Z$ on $r$ can be written down in terms of the dependence of $Z$ on $m_r$ as follows:
\begin{equation}
\frac{dZ}{dr} = 4 \pi r^2 \rho(r) \frac{dZ}{dm_r} \ \  .
\end{equation}

\subsubsection{Mechanical Equilibrium}

We're operating under the assumption of nothing moving... that is, there is mechanical (hydrostatic) equilibrium within the entire stellar model.  This means that the downward force of gravity at any position $r$,
\begin{equation}
 \rho(r) g(r) = \rho(r) \frac{Gm_r}{r^2}
\end{equation}
must be balanced by the {\em}balance of the pressure upwards:
\begin{equation}
P(r)-P(r+dr) = - \left( \frac{dP}{dr} \right) dr.
\end{equation}
If they do indeed balance, then this equilibrium requires that
\begin{equation}
\frac{dP}{dr} = - \frac{Gm_r}{r^2} \rho(r) \ \  \  {\rm or} \ \  \  \frac{dP}{dm_r} = - \frac{Gm_r}{4\pi r^4}.
\label{HSE}
\end{equation}

Note, in passing, that equations \ref{continuity} and \ref{HSE} contain only $r$, $m_r$, $P$, and $\rho$.  So, under circumstances where one can write the pressure in terms of the density only (i.e. without an explicit temperature dependence) then these two equations, plus the $P(\rho)$ relation, suffice to completely describe the mechanical structure of the model.  As an example, consider a polytropic equation of state  like
\begin{equation}
P(r)=K \rho^{\gamma}(r).
\end{equation}
Under this condition, then the two equations can be written in terms of the independent variable ($r$) and two dependent variables ($P$ and $m_r$):
\begin{eqnarray}
\frac{dP}{dr} & = & \frac{Gm_r}{r^2} \frac{1}{K^{1/\gamma}} P^{1/\gamma}(r) \\
\frac{dm_r}{dr} &=& 4 \pi r^2 \frac{1}{K^{1/\gamma}} P^{1/\gamma}(r)
\end{eqnarray}

That is, two equations, two unknowns -- so just add a pair of boundary conditions, and you have a complete stellar model, in hydrostatic equilibrium!

\subsubsection{Energy Generation }

In the more common case where we do care about the thermal content of stellar material, we need to determine the energy balance within each zone of the star.  To this end, we appeal to conservation of energy: that is, the energy flowing into a zone must be balanced by the energy flowing out of that same zone, possibly affected by energy production or consumption within the zone.  If no energy is gained or lost, then clearly
\begin{equation}
L_{r} = L_{r+dr}.
\end{equation}
But more generally,
\begin{equation}
L_{r+dr} = L_r + 4\pi r^2 \rho(r) \left( -\frac{dQ}{dt} + \epsilon \right) dr,
\end{equation}
where $\epsilon$ is the net energy generation rate per unit mass, and $Q$ is the heat content per unit mass.
The quantity $\epsilon$ generally denotes nuclear processes (i.e. energy production via fusion) but also includes energy losses through neutrino emission.  In those cases, $\epsilon$ can be computed as a function of the thermodynamic state and the composition, i.e. $\epsilon (\rho, T, X_i)$.

The term involving the time derivative of $Q$ describes the rate of heat gain or loss, per gram, of material.  Generally, we have
\begin{equation}
\frac{dQ}{dt} = \frac{dE}{dt} - P \frac{\partial V}{\partial t}
             = \frac{dE}{dt} + \frac{P}{\rho^2} \frac{\partial \rho} {\partial t}
\label{econsraw}
\end{equation}
where the first term on the right is the rate of change of internal energy, and the second term accounts for any $PdV$ work done on (or by) the zone.  Equation \ref{econsraw} can be recast using the definition of entropy,
\begin{equation}
\frac{dQ}{dt}= T \frac{\partial S}{\partial t}
\end{equation}
which leads to the more familiar form for the energy conservation equation
\begin{equation}
\frac{dL_r}{dr} = 4 \pi r^2 \rho \left(\epsilon - T \frac{\partial S}{\partial t} \right).
\label{econs}
\end{equation}
We note that the time derivative in equation \ref{econs} is the only place where time explicitly appears in the equations of stellar structure.  More on that later.

\subsubsection{ Energy Transport}

Finally, we require an equation that tells us how temperature changes with position in a stellar model.  To get there, we invoke thermal equilibrium - the assumption that since energy is flowing out of the ``top'' of a given region, energy must also be flowing in from the bottom. We model this transport as a diffusive process driven by the fact that the energy density has to change through the region.  This in turn is a result of the continuously increasing radius of each zone along with any temperature change.  Schematically, we can write the radiant flux $F_r$ as follows:
\begin{center}
$F_r$ = energy-density gradient $\times$ speed $\times$ mean-free-path.
\end{center}
In a less schematic form, we have
\begin{equation}
F_r = - \frac{d}{dr}\left(  \frac{aT^4}{3} \right) \times c \times \lambda
\end{equation}
where $\lambda$ is the mean free path between photon scatterings, and $c$ is the speed of light.  Recognizing that for radiation, the mean free path $\lambda = (\kappa \rho)^{-1}$ where $\kappa$ is the opacity per gram, we can rewrite the above as
\begin{equation}
F_r = \frac{4ac}{3\kappa \rho} T^3 \frac{dT}{dr}
\label{frad}
\end{equation}
for the energy flux carried by radiation.  Multiplying by the surface area of the zone and rearranging yields an expression for the gradient in the temperature when radiative diffusion carries the flux:
\begin{equation}
\frac{dT}{dr} = - \frac{3 \kappa \rho}{16 \pi a c r^2}\frac{L_r}{T^3}
\label{dtdrrad}
\end{equation}

We can also express the temperature gradient in more general terms as a function of the pressure gradient:
\begin{equation}
\frac{dT}{dr} = \frac{dT}{dP} \frac{dP}{dr} = \frac{T}{P} \frac{d \ln T}{d \ln P} \frac{dP}{dr}.
\end{equation}
If we define 
\begin{equation}
\nabla \equiv \frac{d \ln T}{d \ln P}
\end{equation}
then 
\begin{equation}
\frac{dT}{dr} = - \nabla \frac{GM_r}{r^2} \frac{\rho T}{P}.
\label{dtdrgen}
\end{equation}
This is a general expression for the temperature gradient because the mechanism of heat transport is not specified, but is contained in the way $\nabla$ is calculated.  

Equation \ref{dtdrrad}, in fact, can easily be transformed to look like equation \ref{dtdrgen}.
Looking at equations \ref{dtdrrad} and \ref{dtdrgen}, we can define a ``del'' for the case when radiation carries the flux:
\begin{equation}
\nabla_{\rm rad} \equiv \frac{3 \kappa_r}{16 \pi a c } \frac{L_r}{T^4} \frac{P}{G M_r}.
\label{delrad}
\end{equation}
In turn, we can now write the temperature gradient in the model as
\begin{equation}
\frac{dT}{dr} = - \nabla_{\rm rad}
 \frac{G M_r}{r^2} \frac{\rho T}{P}
\end{equation}
for the case where $\nabla$ is determined by radiative diffusion (that is, $\nabla = \nabla_{\rm rad}$).   Later, we will consider other forms of heat transport which can provide values for $\nabla$ under a variety of physical conditions (for example, when the material is convective or conductive).

\subsection{The Constitutive Relations - Where the Physics Is}

The previous section summarized the four differential equations of stellar structure and the general background from where the came.  Each equation relates the dependent variables to the independent variable, but each also includes some other factors and terms, within which the physics that governs stellar structure are reside.  In this section, we outline these quantities and describe the ``constitutive relations'' that provide the route to evaluating the values of these terms.

\subsubsection{The Equation of State}

The dependent variables $P$ and $T$, along with the compositional mix of the stellar material (mass fractions $X_i$ for each atomic species $i$) suffice to determine the density $\rho$, the ionization state of the material the relevant thermal quantities such as internal energy, and the ionization state of the material.  The route to calculating these quantities can be a difficult one in complete generality, but with a few simplifying assumptions we can make progress.  The principal assumption is that the material in a stellar interior is everywhere in local thermodynamic equilibrium, and that all quantities are isotropic.  The fact that the photon mean free path is much smaller than the length scales of interest (the pressure scale height, stellar radius, etc.) ensures that this approximation is a relevant one.

With that assumption (and assuming isotropy) one can show that for an unionized gas, the pressure and density are related through
\begin{equation}
P=\sum_i n_i k T \ \ {\rm where} \ \  n_i=N_A \rho X_i/A_i .
\label{pvnrt}
\end{equation}
We now define the ``mean molecular weight'' $\mu$ as
\begin{equation}
\mu^{-1} \equiv \sum_i X_i/A_i
\end{equation}
so that equation \ref{pvnrt} becomes
\begin{equation}
P = \frac{\rho}{\mu}N_A kT
\end{equation}
and also the internal energy per gram is
\begin{equation}
E= \frac{3}{2} \frac{P}{\rho}.
\end{equation}

Of course, the interiors of stars are mostly ionized, so that the above equations need some modification to account for the free electrons in addition to the nuclei.  Thus we have
\begin{equation}
P = P_e + P_I = P_e + \frac{\rho}{\mu_I} N_A kT
\label{idealeos}
\end{equation}
and further note that the right-hand term for $P_I$ remains valid even in the case of electron degeneracy when dealing with most stars that have asteroseismic potential.  On the other hand, for the electron pressure under non-degenerate conditions, we have
\begin{equation}
P_e = \frac{\rho}{\mu_e} N_A k T
\label{idealPe}
\end{equation}
where $\mu_e$ depends on the ionization state of the material.

As the reader might expect, there are complications beyond simply computing the ionization state.  Departures from ideal gas behavior can be significant under conditions found within the Sun and the stars, and we will discuss those later.  In massive stars, radiation pressure ($P_{\rm rad} = aT^3/3$) becomes increasingly important in the outer layers as the mass increases.

\subsubsection{Energy Generation Rates \label{secengen}}

Stars are, through much of their visible lifetime, gigantic nuclear furnaces.  To ensure energy conservation as per equation \ref{econs} we need an accurate determination of the net energy generation rate per gram, $\epsilon$.  For the nuclear fusion process, this means computing the rate of interaction between a target and a projectile.  We express the probability of these interactions in terms of a cross section $\sigma$ for a given process.  Then the rate of the interaction is proportional to $n \sigma v$, where $n$ is the number density of targets and $v$ is the relative velocity of the reactants.  The velocity will depend on energy (and scale with $\sqrt{kT}$).

The reaction cross section $\sigma$ depends on energy as well, and theory can provide a reasonable functional form of that energy dependence, but the scaling of that relation requires laboratory measurement.  Since the energy dependence of the velocity and cross section can be understood, common practice is to tabulate $\langle \sigma v \rangle$, averaged over the energy distribution corresponding to a temperature $T$.

For resonant reactions, your stellar structure class should have taught you that
\begin{equation}
\langle \sigma v\rangle_{ij} = K_1 \,g \frac{\Gamma_i \Gamma_j}{\Gamma} T^{-3/2} e^{-K_2/T}
\label{res}
\end{equation}
where $K_1$ and $K_2$ are constants that depend on the properties of the interacting particles, and the lifetimes of the incoming state and outgoing state are indicated by the energy widths $\Gamma_i$ and $\Gamma_j$, with the total lifetime given by $1/\Gamma$.  The quantity $g$ is a statistical weight factor.  In the case where a reaction does not have a resonance in the energy range of interest, the $\langle \sigma v \rangle$ value is given instead by
\begin{equation}
\langle \sigma v\rangle_{ij} = \frac{K_0 S(0)}{Z_i Z_j}  T^{-2/3} e^{-K_3 T^{-1/3}}.
\label{nonres}
\end{equation}
Here, $S(0)$ is the ``astrophysical S factor'' which is the value, at zero energy, of a slowly varying function of energy that helps isolate a part of the energy dependence of the reaction.  Once we have determined these cross sections, the rate of energy production per gram of material can be computed with knowledge of the energy yield per reaction and the abundances of projectiles and targets: 
\begin{equation}
\epsilon_{ij}(\rho, T, X)=Q_{ij} \, \rho N_A^2 \frac{X_i}{A_i} \, \frac{X_j}{A_j} \langle \sigma v \rangle_{ij}
\label{engen}
\end{equation}

\subsubsection{Radiation, Conduction, and Convection}

\paragraph{Radiative Transport\\}

Section 1.2.4 expressed the temperature gradient in terms of $\nabla$, which is the local power law slope of the $T(P)$ relation.  The value of $\nabla$ will reflect the dominant energy transport mechanism.  The example shown in that section was for energy transport by photon diffusion, which defined a $\nabla_{\rm rad}$.  That quantity depends in turn on the conditions at that point ($P, T, M_r, L$), some constants, and the radiative opacity $\kappa_r (\rho, T, X_i)$.  The radiative opacity represents scattering photons through a variety of atomic processes involving electronic transitions (bound-free, bound-bond) and free particles (free-free and electron scattering).  Though the scattering cross-section will be a sharp function of frequency (we live in a quantum world) the radiative opacity to be used in computations of isotropic stellar interiors may be simplified using a suitable average opacity over the expected frequency or wavelength distribution of the flux at the local temperature $T$.  A favorite frequency average is the Rosseland mean opacity 
\begin{equation}
\frac{1}{\bar{\kappa}_r} \equiv 
\frac {\int_0^{\infty} \frac{1}{\kappa_{\nu}} \frac{\partial B_{\nu}}{\partial T} d\nu}{\int_0^{\infty}  \frac{\partial B_{\nu}}{\partial T} d\nu}
\label{rosseland}
\end{equation}
where the average is weighted by the local flux derivative (and suitably normalized).  Note that this is an average of the reciprocal of the opacity, since the radiative flux is inversely proportional to the opacity (see equation \ref{frad}) and it is $F$ that is the quantity that is driving the temperature gradient.

\paragraph{Conductive Transport\\}

Under certain circumstances in stellar matter (as well as the much more common household applications), heat can be transported by conduction.  Under the constraints of the formalism adopted in Section 1.2.4, we must seek a {\em conductive} opacity $\kappa_c$ that encapsulates the physics of conduction.  Thus, we write 
\begin{equation}
\kappa_c \equiv \frac{4acT^3}{3\rho D_e} 
\end{equation}
where the diffusion coefficient for conductive heat transport is 
\begin{equation}
D_e \approx \frac{c_v v_e \lambda}{3}.
\label{condif}
\end{equation}
In equation \ref{condif}, $c_v$ is the specific heat at constant density, $v_e$ is the thermal velocity of the electrons, and $\lambda$ is again a mean free path.  Skipping several steps, we can eventually show that
\begin{equation}
\kappa_c \propto \frac{\mu_e^2}{\mu_I} Z_I^2 \left( \frac{T}{\rho} \right) ^2
\end{equation} 
where the constant of proportionality is the hard part.  Finally, we note that when both conductive and radiative opacities are comparable, the combined opacity adds in parallel; that is since 
\begin{equation}
F = F_r + F_c = -\frac{4acT^3}{3 \rho}\frac{dT}{dr} \left( \frac{1}{\kappa_r} + \frac{1}{\kappa_c} \right)
\end{equation} the net opacity is
\begin{equation}
\kappa = \left( \frac{1}{\kappa_r} + \frac{1}{\kappa_c}\right )^{-1}.
\end{equation}

\paragraph{Convection\\}
 
Finally, we come to one of the continuing perplexities of computational stellar evolution, and that is the treatment of the convective flux.  Again, as the energy transport problem has been posed, the problem reduces to determining the value of $\nabla$ when convection carries the flux - that is, when heat transport via bulk turbulent motion, we employ $\nabla_{\rm conv}$ in equation \ref{dtdrgen}.

Convection is required should the radiative temperature gradient exceed the adiabatic temperature gradient.  To demonstrate this, consider an ephemeral blob of material where the temperature exceeds the ambient surrounding temperature.  Such a blob will have a lower density than the surroundings since pressure equilibrium is enforced.  Archimedes taught us that material with a lower density than its surroundings experiences a buoyancy force upwards; this blob begins to rise.  If it does so without losing its identity by exchanging heat with its surroundings, then the temperature will drop as it moves upwards to lower pressures.  The temperature will fall along an adiabat.  The surroundings, however, need not be along the same adiabat; the degree to which the surrounding temperature falls is determined by the local value of $\nabla$ and is presumably $\nabla_r$.  If, after a displacement upwards, the blob finds that it is warmer than its surroundings, it will continue to experience an upward force.  That can only happen if the surrounding temperature drops faster than the adiabatic (blob) value - this occurs when $\nabla_{ad} < \nabla_r$.  If not, and the blob is cooler than its surrounding after an upward displacement ($\nabla_{ad} > \nabla_r$), then the blob's density will become higher than the surroundings, and it will sink back to its equilibrium value.

Eventually, a blob that is buoyantly unstable (and therefore has a higher heat content than its surroundings) will eventually dissolve, resulting in a net transport of heat from the inside of the star to the outside.  This excess heat transport will lower the overall gradient somewhat; a cascade of such blobs therefore can globally affect the temperature gradient in the star. . . if $\nabla_r$ is larger than $\nabla_{ad}$ to start.

This convective instability, therefore, is generated in regions of the star where $\nabla_{r}$ is large when compared to the adiabatic value -- from equation \ref{delrad} that is, where $\kappa$ is large or where $L/m_r$ is large.  In the first case (large $\kappa_r$) we often see the effects of convection in cooler regions of stellar envelopes where partial ionization of a dominant species can drive up the radiative opacity.  In the latter case, if energy generation is confined to a small region, so that the luminosity rises rapidly over a small mass range, then convection can be required to carry all of this locally generated flux.  This is the situation in the cores of stars, and in helium shell burning on the asymptotic giant branch.

In the extreme, convection suppresses the temperature gradient in unstable regions, bringing it down to $\nabla_{ad}$ in the case of perfectly efficient convection.  Of course, modeling this process as resulting in neutrally stratified material is a simplification of the turbulent and dynamic motions expected where convection is present.  Approximating $\nabla_{\rm conv}$ as $\nabla_{ad}$ in regions that are unstable, and mixing the material completely to simulate these dynamical processes, may be good enough for some purposes (such as core convection).  And we have the added benefit that $\nabla_{ad}$ is a thermodynamic quantity, meaning that it comes for ``free'' from the equation of state constitutive relation.  

But in reality we should allow for imperfect convection, and seek a truer treatment of convection that can yield a value for $\nabla$ that is intermediate between $\nabla_r$ and $\nabla_{ad}$.
One such treatment, with a small number of free parameters, is the famous mixing-length theory.  Tuning the mixing length to pressure scale height ratio ($\alpha \equiv \ell/\lambda_P$) so that models reproduce the solar radius and luminosity at the solar age, for example, provides a touchstone to explore how convection might work in other stars.  Many variants of the mixing length have been proposed and employed over the 52+ years since it was introduced into astrophysics by Erika B\"ohm-Vitense \citep{bv58}.

\subsection{Conclusions - room for improvement?}

Given the importance of convective energy transport across stellar astrophysics, and the relatively crude model we have been using for it for decades, it is not surprising that astrophysics is being held back by our ignorance of the convection process.  Computational hydrodynamics simulations of conditions found in stars are being brought to bear on the subject, but are (still) computational expensive and time consuming.  Implementation into stellar evolution codes is a relatively young endeavor, but an important one that is relying on seismic diagnostics of the Sun and stars.

As we will see in the remaining portions of my contribution, the coefficients of the equations of stellar structure rely on basic physics that, in some cases, is well understood.  However, like convection, many of the processes that go into these coefficients are not very tightly constrained by experiment or by theory.  With asteroseismic probes, we may be able to place new and strong constraints on a variety of physics problems.

\section{The physics behind the coefficients}

In the previous section, we reviewed how the basic equations of stellar structure result from various aspects of equilibrium in a self gravitating (mostly) gaseous sphere.  The coefficients within those equations connect with some basic physics: the equation of state of material under extreme conditions, nuclear reaction cross sections, and the interaction between matter and radiation.   In this section we review the relevant constitutive relations and point out areas where the physical inputs could be improved through the indirect probes provided by asteroseismology.

\subsection{The Equation of State: $\rho(P, T, X_i)$ \label{secteos}}

\subsubsection{Basic elements\label{secteosbasic}}
The ``easy'' form of the equation of state involves the assumption of a perfect gas (perhaps undergoing ionization).  Complicating factors include the role of electron degeneracy, but the computational aspects of that are well known.  Less manageable aspects include non-ideal effects, where some (electromagnetic) interaction between particles leads to effects like pressure ionization, Coulomb interactions, and state changes such as crystallization.

To begin with, though, the ideal gas relation leads to an equation of state of the form shown in equations \ref{idealeos} and \ref{idealPe}, where $\mu_e$ depends on the ionization state of the material.  Ionization balance, in the ideal case, is provided by recursive solution of the Saha equation
\begin{equation}
\frac{n_{k+1}}{n_k}n_e = \frac{u_{k+1}(T)}{u_k(T)} \frac{2(2\pi kTm_e)^{3/2}}{h^3} e^{-\chi_k/(kT)},
\label{saha}
\end{equation}
where the index $k$ denotes the ionization level, $\chi_k$ is the ionization potential of level $k$, and $u_k$ is a suitable partition function.  

These all work well enough when the density is ``sufficiently'' low.  But as the density increases, the quantum statistical nature of the electrons begins to overwhelm the simple ideal, perfect gas assumption.  For Fermi--Dirac particles such as electrons (with available spins $\pm 1/2$), the number density is a function of particle momentum:
\begin{equation}
n(p) = \frac{2}{h^3} \frac{1}{\exp{[-\mu + mc^2 + E(p)]/kT}+1}
\label{FD}
\end{equation}
where, generally,
\begin{equation}
E(p) = (p^2 c^2 + m^2 c^4)^{1/2} - mc^2,
\end{equation}
and the chemical potential is $\mu$.

From these, one can obtain the pressure and internal energy from
\begin{equation}
P= \frac{1}{3} \int_p n(p) pv 4\pi p^2 dp \,\,\, - \, {\rm and} - \,
E= \int_p n(p) E(p) 4 \pi p^2 dp \, .
\end{equation}
The interesting part of the above is the energy (and temperature) dependence (the left-had term of equation \ref{FD}, which allows us to write
\begin{equation}
n = \frac{8\pi}{h^3} \int_0^{\infty} F(E) p^2 dp
\end{equation}
If we now define the Fermi energy in terms of the Fermi potential $\mu_F$:
\begin{equation}
E_F = \mu_F -mc^2
\end{equation}
then in the limiting case of low $T$ (that is, $kT \ll E_F$) is $F(E)= 1$ if $E\le E_F$ and $F(E)= 0$ if $E > E_F$.  At these low temperatures, then, $P$ and $E$ become independent of $T$ and depend only on the Fermi energy (which in turn depends only on the density).  This is the well known degeneracy pressure for a $T=0$ system.  At higher temperatures, thermal energy can bump particles to $E > E_F$ at the top of the energy range only, eroding the ``edge'' of the $F(E)$ distribution at the $E=E_F$ edge.  Computation of the electron pressure under this intermediate-degeneracy range is well understood (but does involve some messier integrals).

\subsubsection{Trouble in an ideal world - charged particles {\sl do} interact}{\label{sectinteract}}

In equation \ref{saha}, the relative populations depend on the ionization energies $\chi_k$.  However, those energies are valid for the ion in isolation, and are measured relative to the continuum.  At densities relevant to stellar interiors, however, the electric field from surrounding ions can influence the ionization level by effectively suppressing the continuum, making ionization easier at a given temperature.  The separation between ions, $a$, depends on the density:
\begin{equation}
a= 7.3 \times 10^{19} {\rm cm} \times \left (\rho \frac{X}{A} \right)^{-1/3}
\end{equation}
so that the surrounding ions lead to a depression of the continuum of
\begin{equation}
\Delta \chi \approx \frac{Z^2 e^2}{2a} = 9.8 {\rm eV} \times Z^2 \left(\rho \frac{X}{A} \right) ^{1/3} \, .
\end{equation}
For hydrogen, at densities of a few g/cm$^3$, the effective change in $\chi_H$ is comparable to the ionization energy itself -- that is, hydrogen should be completely ionized independent of the temperature.  It is this ``pressure ionization'' that is responsible for the interiors of stars being fully ionized.  Without it, the high densities within stellar interiors could lead to recombination if the Saha equation remained unmodified -- that is, if the unadulterated ionization potential is used.
Another consequence of these proximity effects is that the internal energy is reduced below perfect gas levels.  Simply, the ion--electron interactions result in a binding energy that would not exist otherwise, adding a negative component to the total internal energy.

If the ions and electrons ``feel'' one another, then at some level, there are ion--ion interactions as well.  This is because the shielding of ions by surrounding electrons can become imperfect if the density is high enough.  Under those circumstances, we must consider Coulomb repulsion between the imperfectly shielded ions.  The energy of these Coulomb effects be proportional to $Z^2 e^2 / a$.  Scaling this to $kT$ gives us the Coulomb parameter $\Gamma_C$:
\begin{equation}
\Gamma_C \equiv \frac{Z^2 e^2}{akT} = 2.27\, \,  Z^2 \left( \frac{\rho}{10^6} \right)^{1/3} 
                    \left( \frac{T}{10^7} \right)^{-1}
                    \left( \frac{A}{12} \right)^{-1/3}
\end{equation}
using cgs units for scaling the density, Kelvin for the temperature, and AMU for the atomic weight $A$.   When $\Gamma_C = 1$, Coulomb effects begin to be significant; in the solar interior, $\Gamma_C=0.1$.  While still small, the sensitivity of helioseismic analysis demands that Coulomb effects be taken into account.  For other stars, such as red giant cores and white dwarfs, $\Gamma_C$ can be much greater than 1, and mutual ion repulsion becomes an important element of the equation of state budget.

In particular, as first proposed by \citet{Sal61}, in white dwarfs where $\Gamma_C$ can exceed 100, this mutual ion repulsion could result in crystallization of white dwarf interiors.  In the modern formulation, the onset of crystallization is believed to occur when $\Gamma_C$ exceeds 175.  In that case, when a white dwarf interior cools to below a threshold temperature
\begin{equation}
T_{\rm xtal} \approx 3.4 \times 10^6 \, \left( \frac{Z}{8} \right )^2 \left( \frac{A}{16} \right)^{-1/3} \left( \frac{\rho}{10^6 {\rm g\, cm^{-3}}} \right)^{1/3} \, K
\label{txtaleq}
\end{equation}
the core solidifies.  Note that the crystallization varies as with $Z^2/A^{-1/3}$, meaning that as a white dwarf cools, heavier elements crystallize at higher temperatures, so in the course of white dwarf cooling, they crystallize {\sl before} lighter elements.  The order of crystallization remains a difficult issue, because of rotation and other mass motions in the degenerate interior.  Whether the core of the white dwarf solidifies as some sort of alloy (principally of oxygen and carbon) or if the oxygen precipitates out prior to reaching the lower temperature of carbon crystallization, remains an interesting and difficult question.

\begin{figure}
\includegraphics[scale=0.37,angle=90]{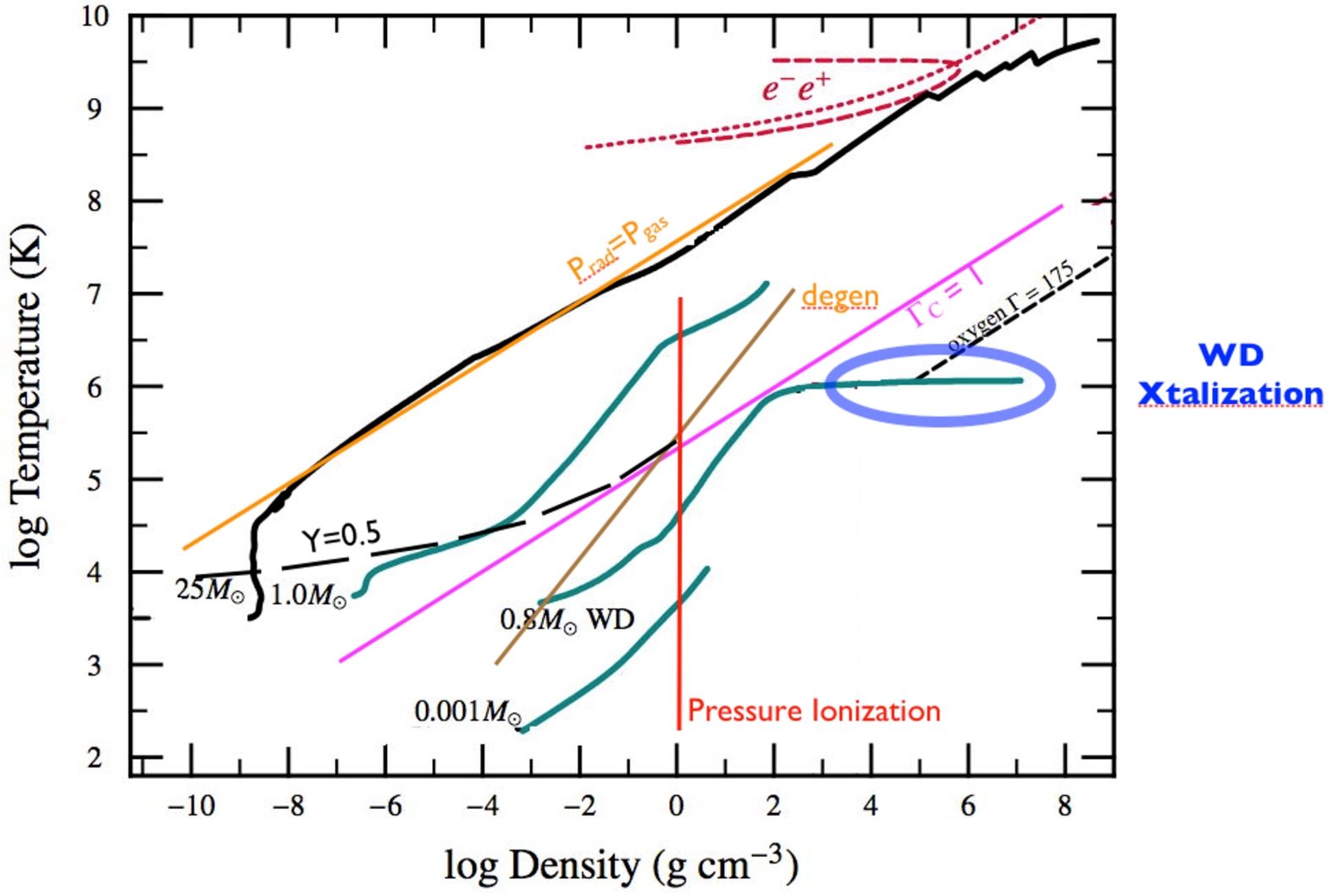}
\caption{Equation of state regions in the $\rho-T$ plane, adapted from \citet{Paxetal11}.  The labeled lines show borders where various effects are relevant as described in the text.  The sold lines also denote the run of density and temperature for stellar models on the main sequence and for a $0.8M_{\odot}$ white dwarf.\label{eosdiag}}
\end{figure}

\subsubsection{Full-up Equation of State Calculations}{\label{sectfulleos}}

Figure \ref{eosdiag} summarizes the various places in phase space where these effects become important.  Putting together the elements of Sections \ref{secteosbasic} and \ref{sectinteract}, the computation of a complete, realistic equation of state is not trivial.  One must address ionization balance using a full mixture of elements and realistic partition functions for the bound states, while also accounting for degeneracy and non--ideal effects.  Most modern treatments use variations of the free-energy minimization approach to compute extensive tables with equation of state parameters for mixtures of interest at a range of pressures (or densities) and temperatures.

As an example, the OPAL effort (see \citep{Rogetal96} and references therein) covers conditions relevant to the Sun and solar-type stars.  For lower temperatures and relevant densities, \citet{Saumetal95} produced widely used tables.  Under conditions relevant to white dwarfs and red giant cores, the most recent useful algorithm (with analytic fits to the full calculations) is by \citet{PotCha10} which allows for Coulomb interactions as well as crystallization.

\subsection{Energy Generation and Loss Rates: $\epsilon(\rho, T, X_i)$ \label{sectnuke}}

\subsubsection{Nuclear cross sections}

Developing expressions for the rate of energy generation through nuclear fusion (and loss through neutrino emission) is at the crossroads between experimental and theoretical nuclear physics and computational astrophysics.  The essential scheme is a collision--physics one: this is a ``rate~=~$n \langle \sigma v \rangle$'' problem at its core.  The relative velocity $v$ depends on the mean kinetic energy per particle ($v(E)=\sqrt(kT/m)$) assuming a Boltzmann distribution for particle velocities, and $\sigma(E)$ is the ``cross section'' for the given interaction.

Nuclear physics provides a functional form for the cross section $\sigma(E)$, but laboratory measurements are essential for an absolute determination for use in stellar models.  Generally, these semi-empirical cross sections are tabulated as $\langle \sigma v\rangle_{ij}$ for interaction between a projectile particle $i$ and target nucleus $j$, as described in section \ref{secengen}.

\begin{figure}
\includegraphics[scale=0.55]{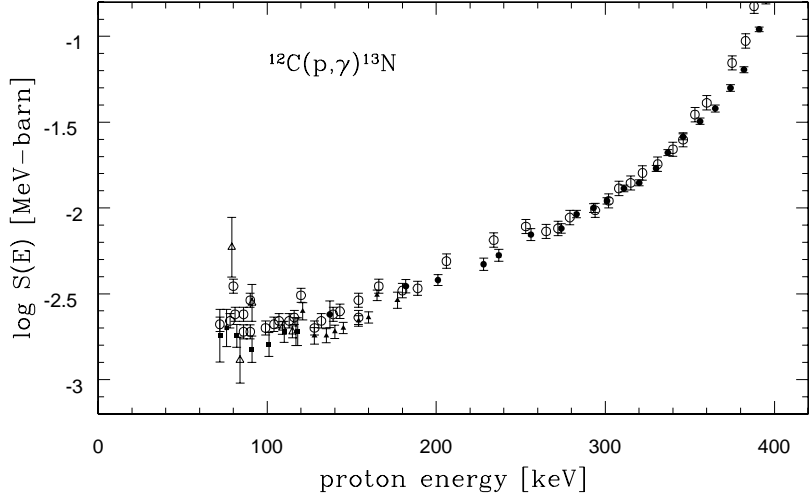}
\caption{Astrophysical $S$ factor for a representative nuclear reaction as measured.  Data from the NACRE compilation (http://pntpm3.ulb.ac.be/Nacre/barre\_database.htm) for the $^{12}C(p,\gamma)^{13}N$ reaction.  Typical relevant stellar energies are approximately 1-2 keV, well below the experimentally accessible energies.\label{sofe}}
\end{figure}

The non-resonant form of $\langle \sigma v\rangle_{ij}$, equation \ref{nonres}, contains the quantity $S(0)$ which is the astrophysical $S$ factor at zero energy.  Why zero  energy?  Simply put, the temperatures that we find in the centers of stars range from 10$^7 K$ for hydrogen burning to a few times 10$^8 K$ for helium burning, or roughly 1-20keV.  Though this represents the mean thermal energy of the particles, it is an order of magnitude or more {\em below} the energies available to nuclear physics experimentation.  In Figure \ref{sofe} we show representative experimental efforts to measure the astrophysical $S$ factor for a nonresonant reaction.  Clearly, the extrapolation to zero energy needs to be done with some care given the experimental uncertainties alone, added to the analytic approximations.  For more precision, additional terms allowing for the value of $S$ at stellar energies can be derived from the experimental data for some reactions.

\begin{figure}

\includegraphics[scale=0.37, angle=90]{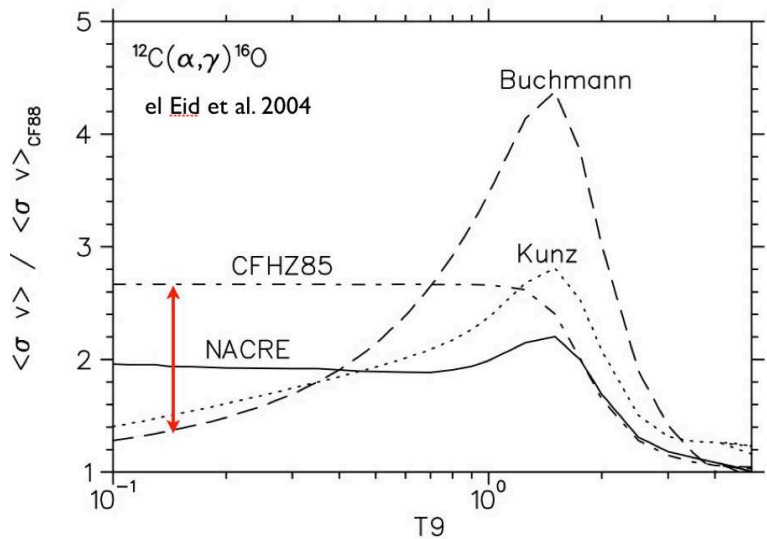}
\caption{Variation of the cross section of the $^{12}C(\alpha,\gamma)^{16}O$ reaction for various formulations, from \citet{eleid04}.  At typical relevant stellar energies are approximately a few $\times 10^8K$ -- indicated by the red line --  the range of suggested values is significant.  See \citet{eleid04} for details.\label{eleid}}
\end{figure}

For {\em resonant} reactions, equation \ref{res} will be effective if the resonance lies within
the range of energies of the interactions (or, more precisely, near the Gamow peak) providing 
that the energy width $\Gamma$, and its position $E_{\rm res}$ is known.  In practice, though, we 
see from Figure \ref{sofe} that the relevant energies are poorly sampled by experiment in many 
cases.  Thus a ``hidden'' resonance can affect the behavior of $S(E)$ significantly, resulting in 
poor estimates of $\langle \sigma v\rangle_{ij}$.  Perhaps the most important example of the 
complicating role of resonances in nuclear fusion rates, in terms of stellar astrophysics, is the 
$^{12}C(\alpha, \gamma)^{16}O$ reaction that is mostly responsible for setting the C/O ratio in 
the Galaxy.  There are low-energy resonances (at about 10 MeV) that are poorly mapped, and affect 
the cross-section at lower energies; see \citet{eleid05} for a discussion.  Depending on the data 
and extrapolation method used, the cross section at temperatures relevant for helium burning (a 
few $\times 10^8 K$) can vary by fairly large factors.  Figure \ref{eleid} shows these cross 
sections for a few formulations of this reaction.

With the above caveats, there are several recent tabulations of nuclear reaction rates and their dependence on temperature and density.  Still, the ``classic'' reference remains \citet{cf88} and references therein, with updated rates available on the NACRE database / website \citep{anguetal99}.

\subsubsection{``Thermal'' neutrino emission \label{sectneut}}

One of the most important coolants for white dwarfs during their early phases is neutrino emission.  The processes that produce neutrinos in white dwarfs differ from the neutrino production associated with nuclear fusion.  In a dense plasma, neutrinos can play the role that photons play in more ordinary stellar material.  When photons are produced in these interactions, they quickly thermalize with the plasma because of electromagnetic scattering.  But when neutrinos are produced, they are not thermalized and leave the star, taking their share of the interaction energy away with them.  Thus neutrino emission is an energy sink, rather than a source.

The ability of neutrinos to act in this way is enhanced when the density is high - and so thermal neutrino processes can be important coolants in the cores of red giants, and in white dwarf stars.  The important thermal neutrino processes include Bremmstrahlung, where neutrinos are involved in free-free scattering instead of photons.  ``Plasmon neutrinos'' are an even more efficient energy sink in hot white dwarfs - they result from decay of plasmons within a dense plasma.  Plasmons are similar to photons, but are coupled with the plasma in such a way that they can decay (into a neutrino -- antineutrino pair) and still conserve momentum and energy.  For an excellent and compact summary of neutrino production mechanisms in white dwarfs, see \cite{wingetal04}.
\begin{figure}
\label{neutl}
\includegraphics[scale=0.55]{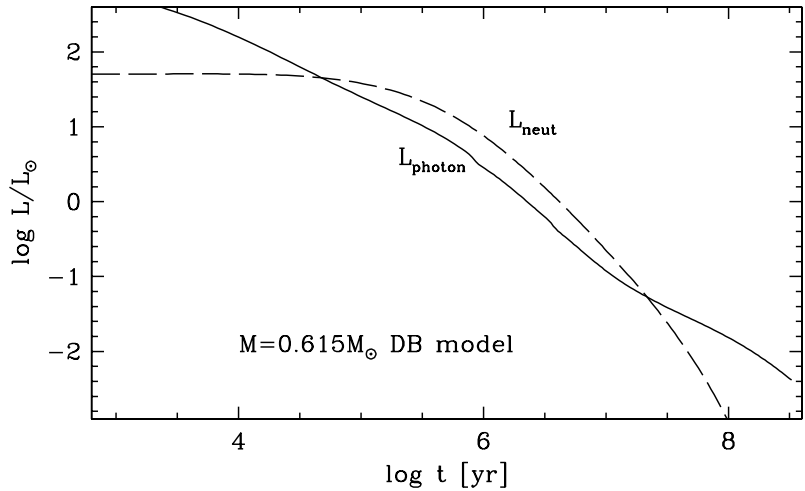}
\caption{Photon and neutrino luminosity as a function of cooling age for a 0.615$M_{\odot}$ white dwarf with a helium--rich atmosphere.  Note that the energy loss by neutrinos exceeds the photon luminosity over a significant time in this early cooling phase.  The plasmon neutrino process dominates Bremmstrahlung for this model.}
\end{figure}

The neutrino emission rates are, as is easy to see, not possible to measure experimentally.  Even if they could be produced under experimental conditions, detection of those produced would be an exciting challenge, since thermal processes do not produce coherent neutrino beams.  The rates are calculated using the Standard Model of particle physics.  A current algorithm for using those calculations in the form of energy loss rates in stars is \citet{itoh96}. Verifying those computations does fall into the realm of astrophysics - neutrinos are the dominant coolant for most white dwarfs during a significant fraction of their early cooling stages.  So, measuring the cooling rate of white dwarfs can provide an experimental, if indirect, test of the theoretical thermal neutrino production rates.  This measurement can be done in a statistical way by looking for features in the luminosity function of a collection of white dwarfs, but a more direct way is to measure the cooling rate of a single white dwarf. Figure \ref{neutl} gives a preview of this, showing that the neutrio luminosity can exceed the photon luminosity during a portion of the cooling history of a white dwarf. We will discuss this further in a later section, but clearly a sensitive asteroseismic probe of white dwarf structure can enable this measurement.

\subsection{Opacities: $\kappa(\rho, T, X_i)$ \label{sectkappa}}

For regions of the star that are not convectively unstable, reliable calculation of the opacity can be non-trivial, especially in the outer layers where the radiative opacity is dominated by atomic processes.  The Rosseland mean opacity (equation \ref{rosseland}) hides these difficulties in the frequency dependent opacity $\kappa_\nu$.  These processes include electron scattering (which is easy as it is not frequency dependent), free--free scattering, bound--free absorption, and bound--bound absorption.  At cooler temperatures $H^-$ opacity becomes significant for solar-type stars, and even cooler stars require accurate treatment of molecular absorption by CO, OH, H$_2$O, CH$_4$, etc.
\begin{figure}

\includegraphics[scale=0.40, angle=90]{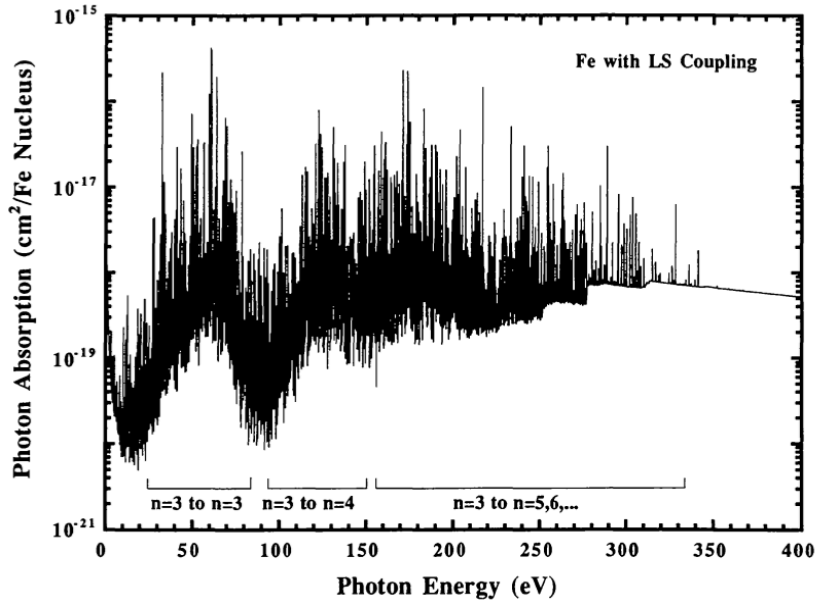}
\caption{Frequency--dependent absorption coefficient for iron at $\log T=5.4$, from \citet{rogig92}\label{kapnuopal}}
\end{figure}
An issue in computing the Rosseland mean opacity is that the frequency--dependent cross section is an extremely rapidly varying function at temperatures near and below complete ionization.  Figure \ref{kapnuopal} illustrates this for iron, from \citet{rogig92}, at $2.51\times 10^5K$, with the relevant atomic transitions as labeled.  

Considering the fact that one needs to include all species that contribute to the opacities, integrals for the Rosseland mean opacity must cover dozens of elements (each in their appropriate ionization and excitation states).  Indeed, advances in modeling atomic states and in computational scale resulted in significant revision of the atomic opacities used in stellar interiors calculations in the mid 1990s by \citet{rogig92} and \citet{seatetal94}.  It is notable that the need for updated opacities (particularly the contribution of iron) was pointed out in part as the result of seismic modeling of Cepheid variables \citep{simon82, iglrog91} as we will see in a later section.

\begin{figure}

\includegraphics[scale=0.40, angle=90]{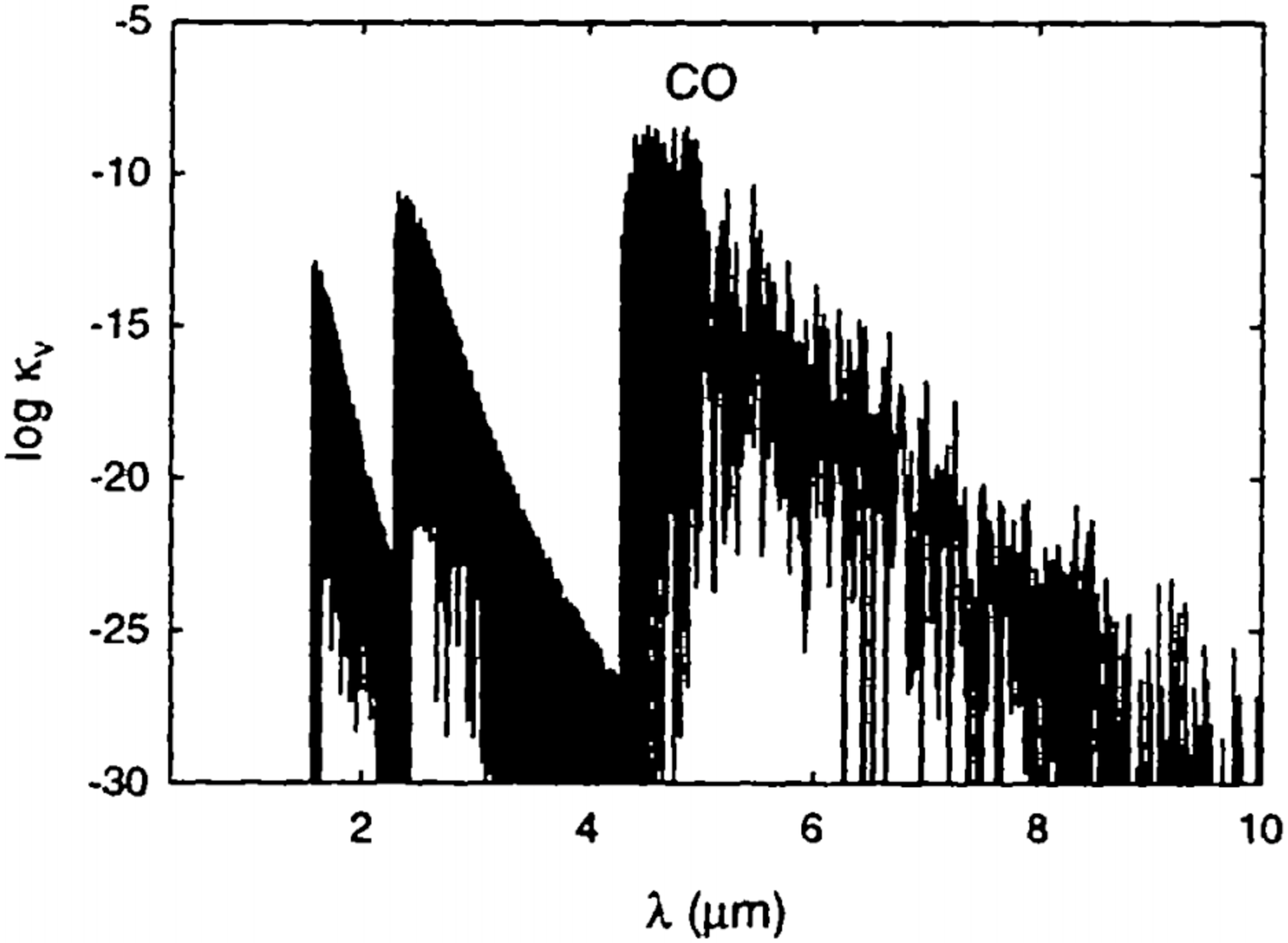}
\caption{The absorption coefficient for CO as a function of wavelength, from \citet{alexetal03}.\label{coopac}}
\end{figure}

For cooler stars, the contribution of molecular transitions (and, at lower temperatures, dust opacity) complicates the calculation further. Figure \ref{coopac}, from \citet{alexetal03}, shows a sample run of the absorption coefficient (per molecule) for CO.
The relative strengths of the various low--temperature opacity sources are summarized by \citet{fergetal05}.

\subsection{Summary}
Overall, stellar models do a very good job of describing many of the observables of stars with the current state--of--the--art of physics inputs (as long as we continue to not mention convection).  When asteroseismic efforts are introduced, though, gaps in our physics understanding are quickly exposed, and this in turn can shake our confidence in the way we treat the input physics for stellar models.  The current section reviewed the approach taken in modern stellar modeling -- and in various places, improvements have been made based on seismic analysis.

In the next sections, we will show how observations of pulsating white dwarfs can test or constrain the physics of dense matter under conditions when crystallization might occur.  Those pulsating white dwarfs also place constraints on interesting nuclear cross sections (in particular, for the $^{12}C(\alpha, \gamma)^{16}O$ reaction, and on the neutrino emission rates.  Cepheid pulsation systematics demanded recalculation of radiative opacities that include a realistic treatment of heavy elements.  Observations of pulsating sdB stars probe other effects such as radiative levitation (as we will see), and solar-like oscillations of main sequence stars, as well as white dwarfs, provide new probes of convection in stars that may finally shake our reliance on the antiquated mixing length theory.

\section[Seismology to the rescue]{Seismology to the rescue: feedback between pulsation studies and input physics}

In the previous section we reviewed some of the more difficult or problematic aspects of the physics of stellar interiors and how that physics can be (compactly0 implemented in stellar models.  Now we can discuss how asteroseismic studies have impacted this area of stellar astrophysics (mostly for the better).  The asteroseismic signal of greatest utility is the set of pulsation frequencies -- using relatively simple adiabatic pulsation theory provides the tool we can apply.  We'll cover some examples without trying to be exhaustive, drawing heavily on results from compact pulsating stars like white dwarfs -- since that is an area that has been around for a while and one that I'm familiar with.  \citet{altetal10} provide a comprehensive review white dwarf evolution in the context of asteroseismology that is complementary to the review by \citet{kawsaas95}.

We start with white dwarf asteroseismology and how we can probe the crystallization of stellar interiors, and also how seismic probes of the internal composition profile can constrain important nuclear cross sections.  Next we'll briefly discuss how timing of white dwarf pulsations over many years can provide a measurement of the cooling rate of individual stars, and therefore constrain neutrino emission rates.  For variety we'll then consider how the theory of Cepheid pulsations led to a revision of the radiative opacity calculations.

\subsection{White dwarf crystallization revealed by asteroseismology}

In Section \ref{sectinteract} we told the story of how \citet{Sal61} concluded that under conditions found within white dwarf stars, the interior could undergo a phase transition and develop a crystalline core.  What \citet{Sal61} did not anticipate was that we would find that white dwarfs undergo nonradial pulsation when the (pure hydrogen) envelopes reach temperatures where hydrogen is partially ionized at depth.  

Generally at $T_{\rm eff}\approx 11,000K$ to $13,000K$ the envelope partial ionization drives convection and also pulsations, through processes to be reviewed later.  The historical development of white dwarf asteroseismology, which dates from the discovery of the first pulsating white dwarf by Arlo Landolt in the late 1960s \citep{land68}; for an excellent overview of the current state of the subject, see \citet{winkep08} and \citet{fonbra08}.
There is a correlation between the luminosity of a white dwarf and the core temperature that follows closely the seminal work by \citet{Mestel} and updated by \citet{VanHorn}:
\begin{equation}
\frac{L}{L_{\odot}} = 1.7 \times 10^{-3} \left (\frac{M}{M_{\odot}} \right) \frac{\mu}{\mu_e^2} T_{\rm c ,7}^{3.5}
\end{equation}
where the core temperature is in units of $10^7$K.  Recalling Section \ref{sectinteract} and in particular equation \ref{txtaleq}, the central temperature at crystallization for oxygen is approximately $3.4\times 10^6$K, which is at a luminosity of only $\approx 2\times 10^{-5} L_{\odot}$.  This is at the lower limit of observed white dwarf luminosities, suggesting that any white dwarfs that are cool enough to crystalize will be too cool to pulsate.  However, one must remember that the radius of a white dwarf is determined almost entirely by its mass (the famous mass-radius relationship for degenerate stars) - so if the mass of a white dwarf at that luminosity is large enough, its would still require a sufficiently high $T_{\rm eff}$ to be seen at this luminosity.

\begin{figure}

\includegraphics[scale=0.40]{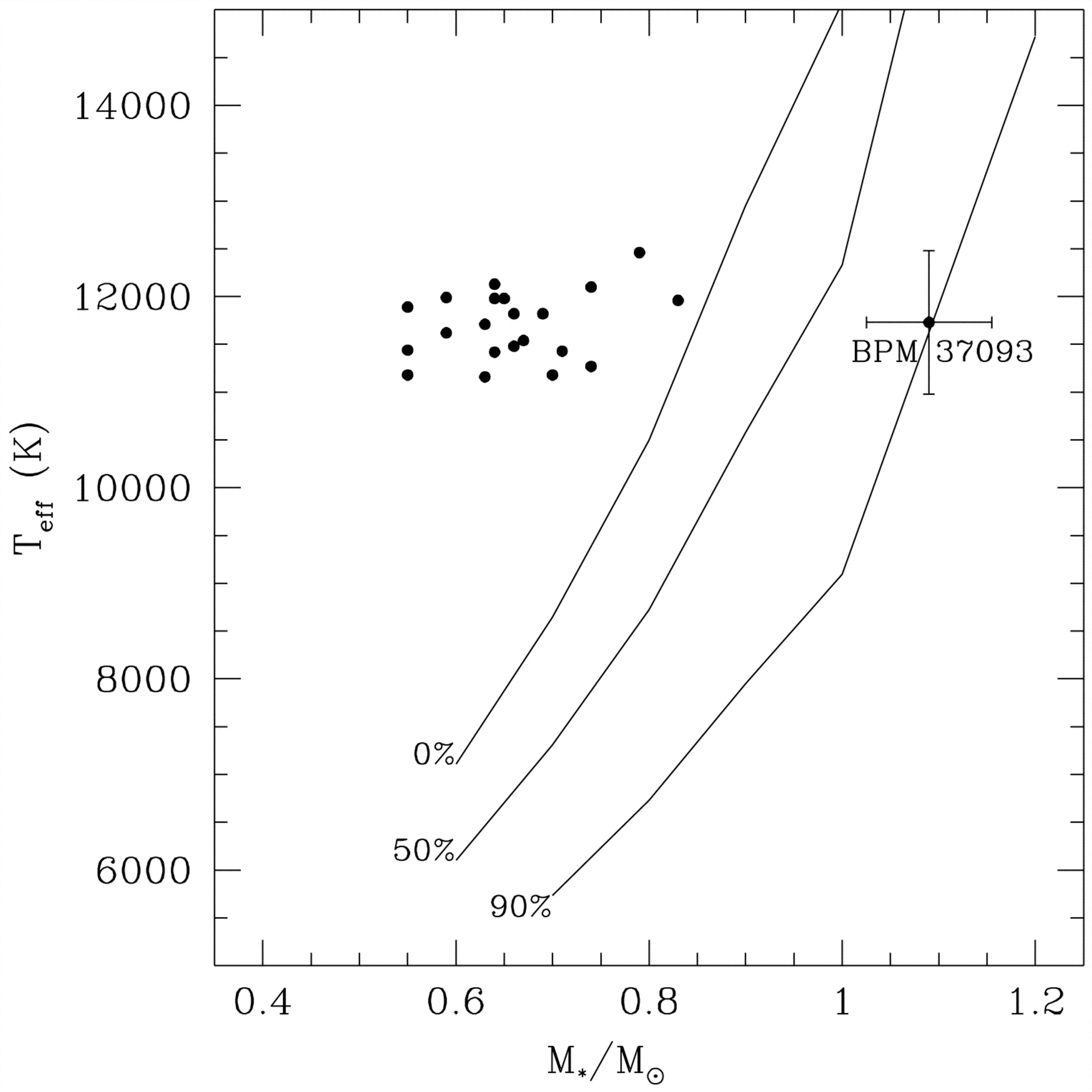}
\caption{Crystalline core fractions for white dwarf models as a function of mass and $T_{\rm eff}$ compared with some pulsating white dwarfs, from \citet{monwin99}.\label{monwinmod}}
\end{figure}

In fact, white dwarfs with masses greater than about 1$M_{\odot}$ can crystalize at effective temperatures that place them in the pulsational instability strip.  This possibility was first explored by Mike Montgomery and Don Winget \citep{monwin99} following the discovery of a massive pulsating white dwarf, BPM~37093, with a mass of approximately $1 M_{\odot}$ \citep{kanetal92}.  Computations of evolutionary white dwarf models by \citet{monwin99} (see Figure \ref{monwinmod}) compared with the then-known pulsating white dwarfs confirmed the theoretical expectation that BPM~37093 should be a pulsating white dwarf with a crystalline core.
The seismic diagnostic of a crystalline core is the fact that nonradial modes propagate quite differently through solids than through gasseous material.  For $g-$modes such as we see in white dwarfs, the eigenfunctions carry very small amplitude in crystalline material when compared to non-crystalline white dwarf cores, which in turn affects the pulsation frequencies in a measurable way.  \citet{monwin99} demonstrate how a of $g-$mode overtones can determine the mean period spacing, and coupled with other data can reveal whether the core is crystalline or not.

\begin{figure}

\includegraphics[scale=0.40]{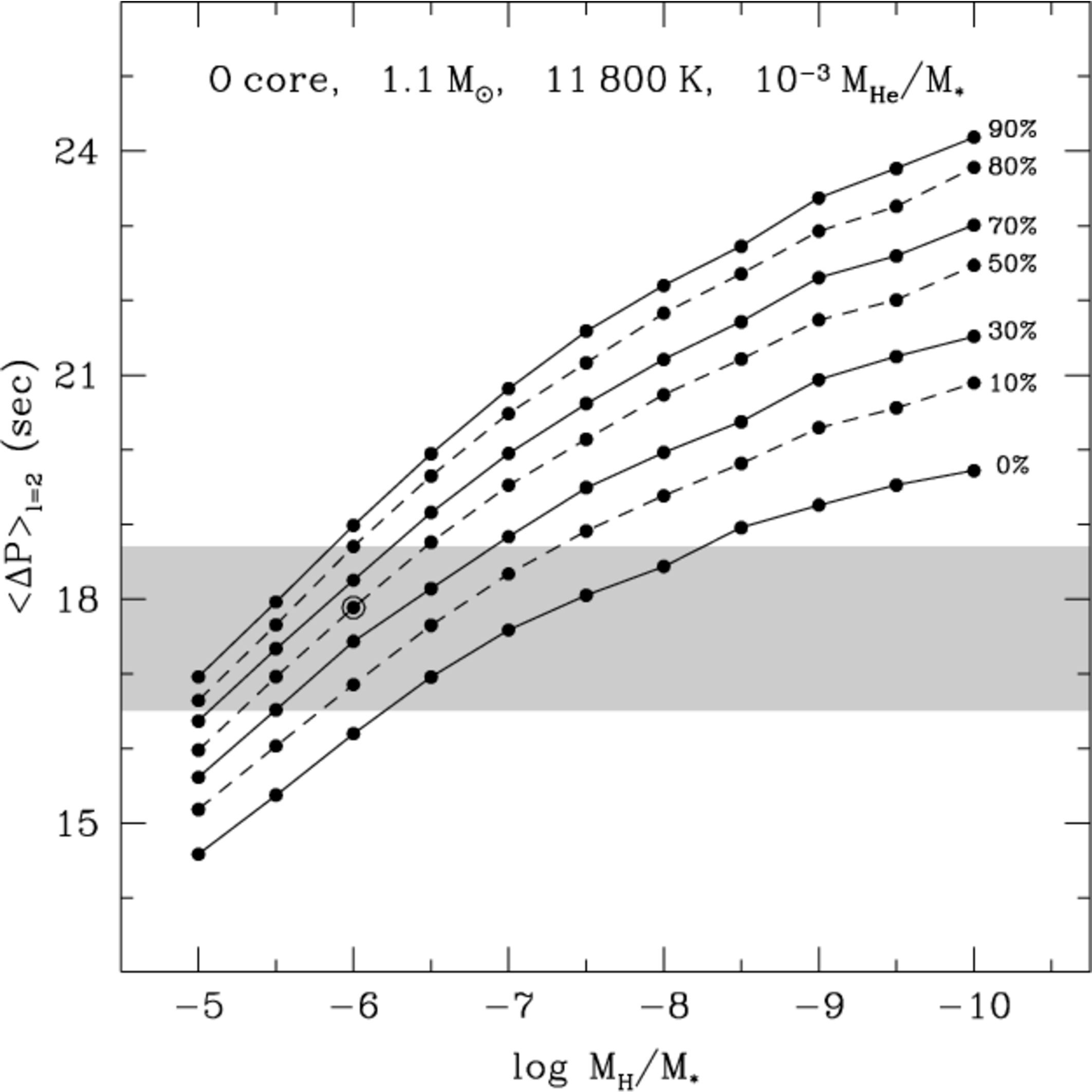}
\caption{The observed period spacing in BPM~37093 compared with relevant stellar models, from \citet{kanetal05}. \label{kanfig3}}
\end{figure}

In an effort to determine the frequencies of as many modes as possible, BPM~37093 was the primary target for two {\it Whole Earth Telescope} campaigns that ultimately were able to expose several $l$=2 $g-$modes in the star \citep{kanetal00, kanetal05}.  Figure \ref{kanfig3}, from \citet{kanetal05} shows the measured period spacing superimposed over various models.  Modeling of these pulsations concluded that there is strong evidence that the core of BPM~37093 is indeed crystalized, in accordance with theoretical expectations.  
With the conclusion that BPM~37093 was largely crystalline, the public became highly engaged.  This asteroseismic result was announced in February 2004, making the discovery of an enormous ``diamond star'' a terrific Valentine's Day story!

\subsection{The $^{12}$C($\alpha, \gamma$)$^{16}$O reaction rate - constraints from white dwarf seismology}

We saw in Section \ref{sectnuke} that the reaction that is principally responsible for the C/O ratio in the Universe has large uncertainties associated with it because of the difficulty mapping low--energy resonances.  The $^{12}$C($\alpha, \gamma$)$^{16}$O reaction occurs at slightly higher temperatures than the triple$-\alpha$ reaction that produces $^{12}$C during core helium burning, and so the rate of production of $^{16}$O increases towards the end of helium core burning.  White dwarfs therefore should have a decreasing C/O ratio as one progresses from the outer core to the inner core reflecting this evolution.  The central C/O ratio will therefore depend on the $^{12}$C($\alpha, \gamma$)$^{16}$O reaction rate.

White dwarf pulsation periods principally depend on the envelope structure \citep{winkep08}, but the core properties also influence the periods.  In matching observed periods with model periods,  near--surface composition gradients (accentuated by diffusion processes) produce mode trapping, allows determination of the surface layering structure.  But composition gradients in the core (caused by the increasing production of $^{16}$O, for example) are also influential \citep{metmonkaw03, monmetwin03}.

\begin{figure}
\includegraphics[scale=0.40]{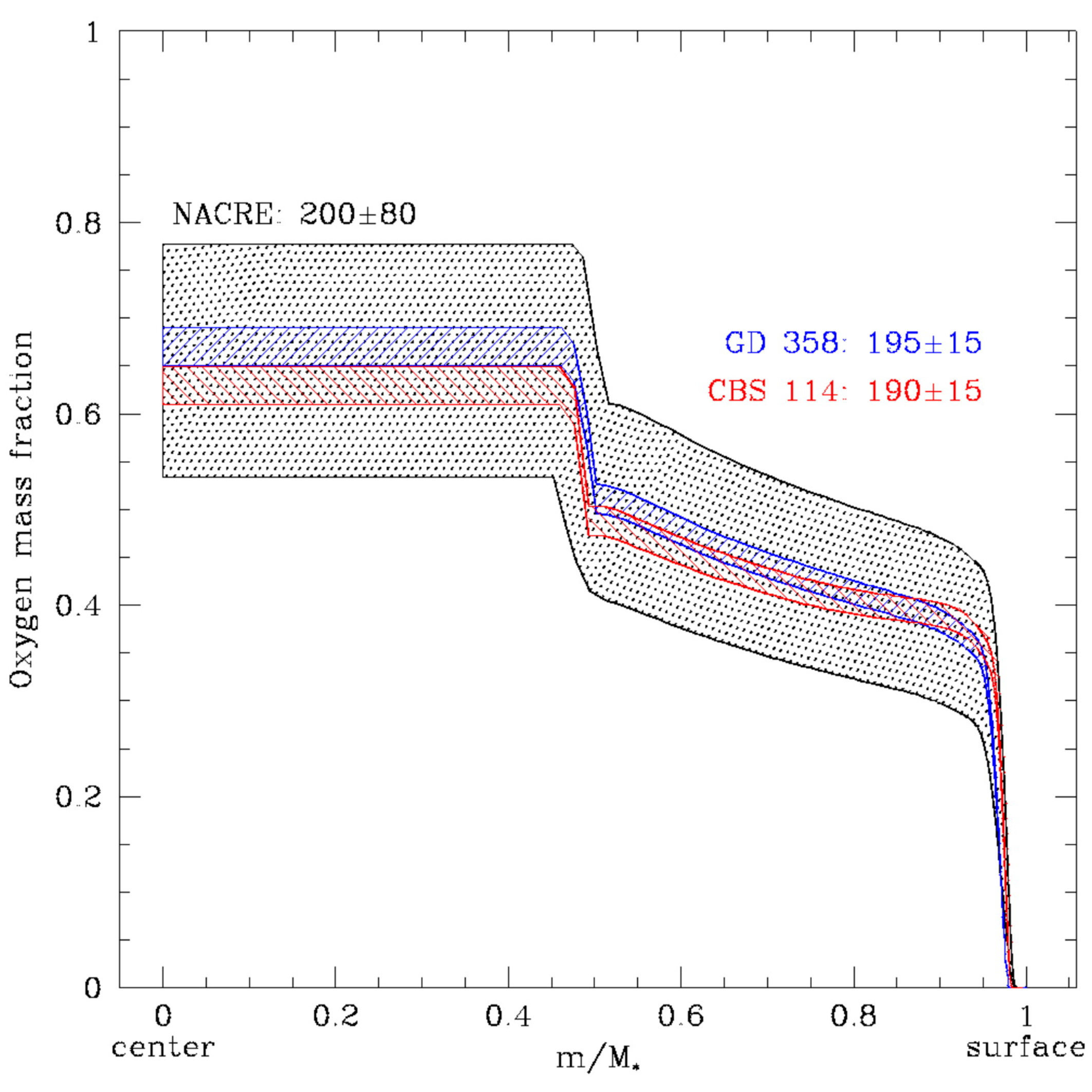}
\caption{Oxygen abundance profile in a pulsating white dwarf interior, from \citet{met03}.  The wide (gray) shaded region is the range of profiles given the $\pm 1\sigma$ range from the NACRE compilation \citep{anguetal99}; the seismic constraints are given in the narrower in band for two white dwarf pulsators.\label{met03fig2}}
\end{figure}

With several DB (helium atmosphere) white dwarfs having successful ground based observing campaigns that revealed a large number of modes, several groups have investigated whether seismic probes could in fact measure the core oxygen abundance and the composition gradients in the core caused by the sequential stages of helium burning.  \citet{metetal01,metetal02} showed that by varying the assumed $^{12}$C($\alpha, \gamma$)$^{16}$O reaction rate, the core oxygen mass fraction in the remnant white dwarfs could change significantly.  Using data from GD~358 \citep{wingetal94}, and CBS~114 \citep{handetal02}, \citet{met03} concluded that the rate for this problematic reaction are in fact in line with recent determinations discussed by \citet{eleid04}.  The seismic results are shown in Figure \ref{met03fig2}.

\subsection{Measuring white dwarf cooling via asteroseismology}

In section \ref{sectneut} and in particular Figure \ref{neutl}, we saw that the cooling (and therefore overall evolution) of white dwarfs can be dominated by plasmon neutrino emission during the drop in luminosity from the planetary nebula nucleus phase into the low-luminosity phase.  By measuring the cooling of an individual white dwarf, whose global properties are otherwise constrained by spectroscopy and asteroseismology, we can in principle test the theoretically expected rate via observation.

\subsubsection{Measuring cooling rates via pulsations}

Consider the accuracy to which we can measure most stellar properties through spectroscopy.  On a good day, we can measure, for example, $T_{\rm eff}$ to 0.1\%.  With knowledge of the distance to a star, we might gain a precision in luminosity of perhaps 1\%.  Now, if we want to measure {\em changes} in those quantities brought about by the slow secular evolution of the star, which occurs on time scales of millions or billions of years, we'd have to stick around for a thousand years or more before being able to detect any changes at all.

But through time-series photometry or spectroscopy, we can measure oscillation frequencies to a much higher precision, especially in those stars that show self-excited (and therefore extremely coherent) pulsations.  Stars for which we can get parts-per-million accuracy in periods include the pulsating white dwarfs, $\delta$ Scuti stars, and other classical pulsators such as RR Lyr and Cepheid pulsators.  With measurement of a global property of a star (a pulsation period) at that level of accuracy, we can indeed hope to measure a change in the pulsation period, caused by secular evolution, in a relatively short period of time (within the career of a graduate student, for example).

The direct measurement of a pulsation period change requires patience and a cooperative star that shows tractable phase or amplitude variations over many years.  While one could in principle measure the pulsation period itself directly at two widely--spaced epochs, a more integrated method would be to look for continuous phase changes caused by the increasing or decreasing period.  If we measure the period very precisely, and use it to predict the phase in the future under the assumption that the star's period is not changing with time, then any real changes in the period within the star will cause the phase to advance (or retreat) in a way that will reveal the period change.  The traditional method for accessing these phase changes is through a so-called $(O-C)$ analysis, where $(O-C)$ is the difference between the observed phase at a given time, and that calculated assuming that the pulsation period is constant.

The phase, as determined from the time of next maximum in the light curve at a given epoch, is given by
\begin{equation}
\label{tmax}
t_{\rm max, i} = t_{\rm max,0} + i \times P_{0} + i^2 P \frac{dP}{dt}
\end{equation}
for the $i$th maximum after an agreed--upon first maximum, where we have expanded the equation through a Taylor expansion assuming $dP/dt$ is small.  If the period itself does not change with time, the predicted maximum of the $i$th cycle will be perfectly accurate with only the first two terms in the above equation, and the value of $(O-C)_i$ will be a constant at zero.  If there is an error in the period, then $(O-C)$ will increase linearly with time (if the assumed period is too short) or decrease linearly with time (if too long).

\begin{figure}
\includegraphics[scale=0.40,angle=90]{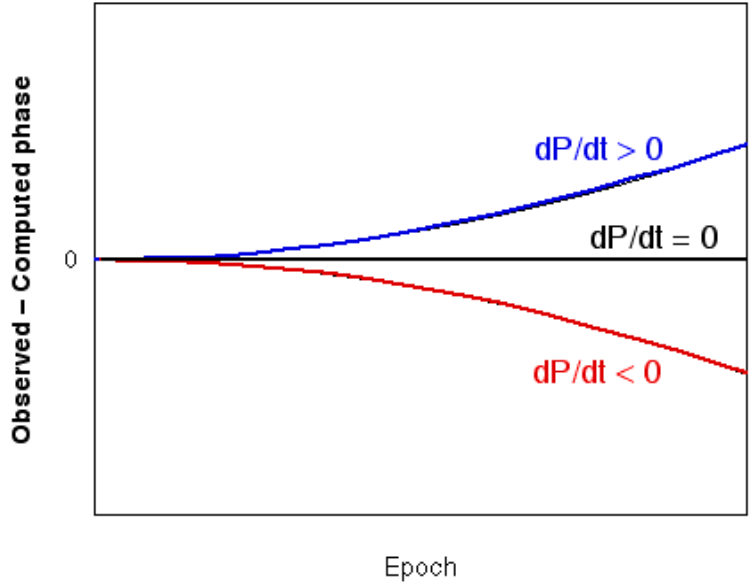}
\caption{The change in $(O-C)$ as a function of time for constant rate of increase in period (upper curve) or decrease in period (lower curve).\label{omcsketch}}
\end{figure}

If the $P_0$ is not in error but the period is changing (slowly) with time, then equation \ref{tmax} yields
\begin{equation}
\label{omc}
(O-C) \ =\  (\Delta t)^2 \frac{1}{P} \frac{dP}{dt} \ \ ,
\end{equation}
where $\Delta t$ is the time elapsed between the chosen epoch and the measurement of the current time of maximum $t_{\rm max}$.  Clearly, the departure of $(O-C)$ from zero increases as (time)$^2$ goes by, enabling quadratically increasing sensitivity to $dP/dt$ over time.  A change in $(O-C)$ will have a parabolic form if $dP/dt$ is constant, as shown in Figure \ref{omcsketch}

Since the pulsation period is some measure of the global structure of the star (such as the radius as it affects the mean density), the period of pulsating stars much change as they evolve, and must change on approximately the evolutionary time scale.  Assuming a a time scale for period change (inversely proportional to $\dot P / P$) of 10$^7$ years, and a precision of measuring $t_{\rm max}$ of 10 seconds (both typical for young, hot white dwarfs), one finds that the time needed to detect an evolutionary change is only about two years for a 1$\sigma$ result, and 3 years for a 3$\sigma$ result.  

Of course, these expectations are for ideal circumstances. For the above process to work, we need to ensure that the frequencies are precisely known, and that the analysis is not hampered by cycle-counting errors during the inevitable daily, monthly, and seasonal gaps in the data.  In addition, not all oscillation modes are not as stable in phase and/or amplitude as needed to reveal this effect.  Nonlinear interactions between modes, shorter-timescale phenomena such as convection or magnetic fields, and other effects can hamper our ability to do the necessary high--precision cycle-counting required.  All of these effects would only lead to larger values for $\dot P$, so that the estimates for the rate of period change (and therefore the evolutionary time scale) are always lower limits to the true time scales of evolution.

White dwarfs evolve at roughly constant radius, so their evolution is a cooling process, with the fading over time resulting entirely from a decreasing core (and envelope) temperature.  The pulsations are $g-$modes, which are sensitive to the Br\"unt-V\"ais\"ala frequency.  That, in turn, is set by the temperature stratification in the star - hence the cooling of the star leads to a decrease in the  Br\"unt-V\"ais\"ala frequency, and an increase in the pulsation period.  Thus for white dwarfs we expect to see periods increase with time, on time scales equal to the cooling rates of those stars.

\begin{figure}
\includegraphics[scale=0.45, angle=90]{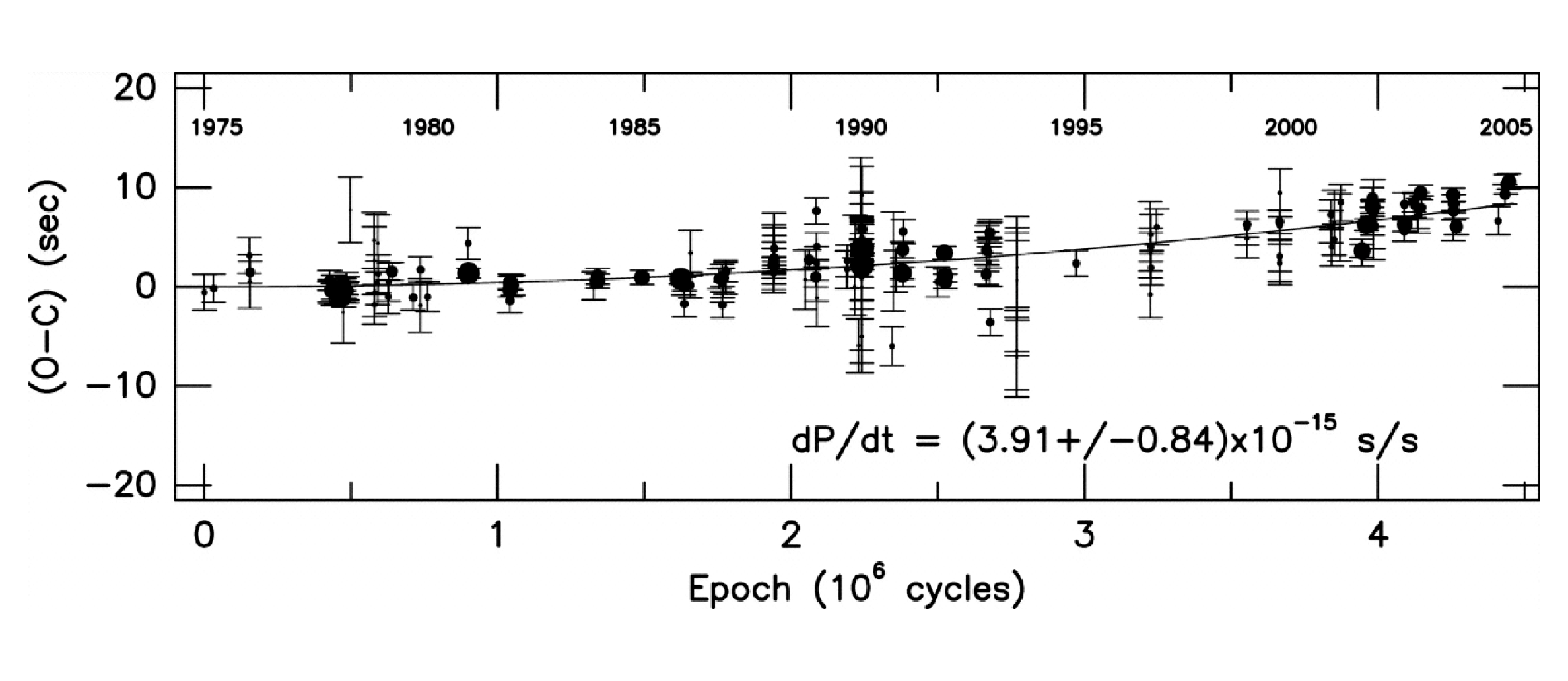}
\caption{An $(O-C)$ diagram for the 215s $g-$mode in the pulsating DA white dwarf G117-B15A, revealing the upward parabola that results from a positive rate of period change, from \citet{kepg117}.  The time scale for period change is approximately 1.7$\times 10^9$ years.\label{g117b15a}}
\end{figure}

An excellent demonstration of this is the pulsating DA white dwarf G117-B15A. This star has been continually monitored for well over 30 years, mostly by the Brazilian astronomer S. O. Kepler (no relation).  Figure \ref{g117b15a} shows the $(O-C)$ diagram for this star, and the (now) statistically significant measurement of the rate of cooling of this star \citep{kepg117}.  This star is significantly cooler than the example cited above, and has an expected cooling time scale of approximately 10$^9$ years.

\subsubsection{Application to white dwarf pulsators}

There are two classes of pulsating white dwarf that overlap with this neutrino cooling phase.  At the higher--luminosity end are the GW~Vir stars, with the prototype PG~1159-035.  The luminosities of those stars are generally higher than of interest for neutrino cooling studies, but the lower--luminosity members are in the range where $L_{\nu} > L_{\gamma}$.
In particular, \citet{obretal98} and \citet{obrkaw00} show that the GW Vir star PG~0122+200 would be a good candidate for measuring the cooling rate and therefore constraining neutrino emission.   Unfortunately, a recent $(O-C)$ analysis suggests that the period of PG~0122+200 is changing much faster than the models predict \citep{vaucetal11}, suggesting that other factors are influencing the pulsations.

\begin{figure}
\includegraphics[scale=0.37, angle=90]{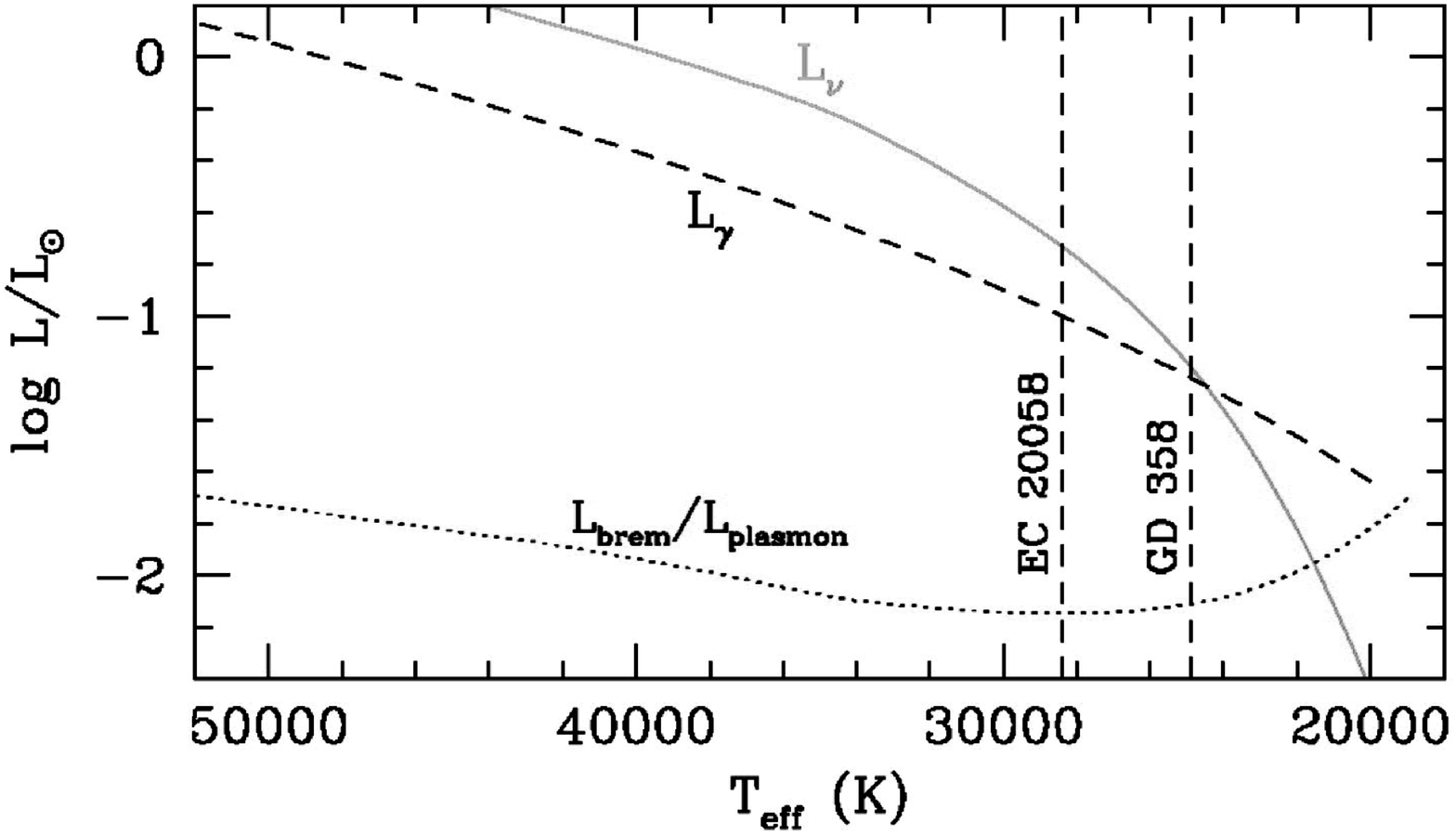}
\caption{Relative photon and neutrino luminosities in the region where we find DB pulsating white dwarfs, from \citet{wingetal04}.\label{neutldb}}
\end{figure}
\begin{figure}
\includegraphics[scale=0.55]{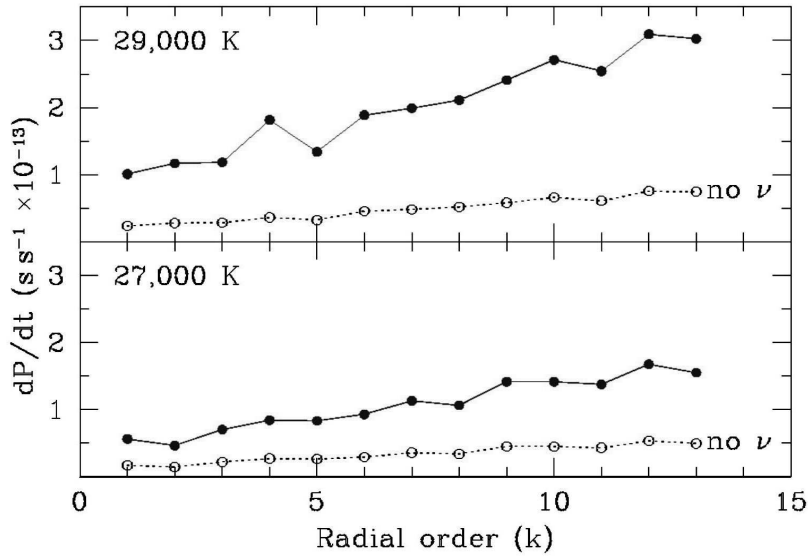}
\caption{Rate of period change in various DB white dwarf models, from \citet{wingetal04}.  The solid curve connects modes in a model with the standard neutrino emission, while the dotted line is for modes in a model with neutrino emission suppressed.\label{dpdtdb}}
\end{figure}

\citet{wingetal04} show an example of this approach (see Figure \ref{dpdtdb}), and demonstrate that the rate of period change can be an effective probe, since the rate of period change depends nearly linearly on the neutrino rates used in the calculations.

More promising are stars at the lower end of the range of luminosities where neutrino emission dominates photon cooling (see Figure \ref{neutl}).  Expanding the lower luminosity side in Figure \ref{neutldb} (from \citet{wingetal04}) we see that two well-studied pulsating DB white dwarf stars straddle the luminosity where neutrino emission yields to photon emission as the dominant cooling process.

In particular, EC~20058 should be experiencing neutrino cooling that exceeds photon cooling by nearly a factor of two.  The influence of neutrinos on the cooling rate can be estimated by computing evolutionary models of the stars with the current neutrino rates, and with modified neutrino rates.  Self-consistent calculations with and without neutrino emission, for example, both constrained to match the observed pulsation periods themselves, can then be probed to provide the expected rate of period change as a function of the neutrino emission rate.  

\citet{suletal08} report on {\it Whole Earth Telescope} observations of EC 20058, which is part of a long--term observing campaign that spans the time from its discovery as a pulsator in the mid 1990s through the present.  The star is an extremely stable pulsator (in terms of phase and amplitude changes) and should, in the near future, provide the first strong constraints on the rate of plasmon neutrino emission from a dense plasma.  As Denis Sullivan would say, ``stay tuned!''

\subsection{Cepheid masses as opacity probes}

Here, I will briefly recount a saga from the annals of stellar pulsation theory that was an early demonstration of the power of asteroseismology (though it was yet to be named that) to expose and fix problems with stellar interior physics.  The centerpieces of this story are multimode pulsators: the so--called ``beat Cephieds'' and ``bump Cepheids.''  The beat Cepheids (also sometimes called double-mode Cepheids) pulsate in the (radial) fundamental and first--overtone modes.  The period range of the beat Cepheids is in the 2-7 day range, and the ratio of the period of the first overtone to the fundamental ($P_{10}$) ranges from about 0.695 (at the long period end) to 0.715 at the short period end.  The ``bump Cepheids'' display a secondary maximum that progresses slowly in phase with respect to the fundamental pulsation mode; the bump is the manifestation of a near--resonance between the pulsation of the second overtone $P_2$ and the fundamental $P_0$.  The bump Cepheids are a more homogeneous class with periods close to 10 days, and $P_{20}$ very nearly 0.5 -- that is, close to a 1:2 resonance.

As summarized very nicely in \citet{mosetal92} and \citet{simon87}, evolutionary models of beat Cepheids with the proper fundamental period $P_0$ had period ratios $P_{10}$ that were significantly larger than the observed value.  To reach the observed $P_{10}$ models using the then-standard opacities (i.e. in the early 1980s) required unrealistically low masses; even so, the models had a steeper dependence of $P_{10}$ on $P_{0}$ than what is observed.  For bump Cepheids, the period of 10 days corresponded to a mass of 6.3$M_{\odot}$ -- but then the theoretical value of $P_{20}$ was too high to explain the observed resonance.  Thus as of 1981, the evolutionary model masses for Cepheids did not match the pulsation masses, with the discrepancy being large enough (a factors of two more) to make this a ``famous problem.''  

\begin{figure}
\includegraphics[scale=0.25]{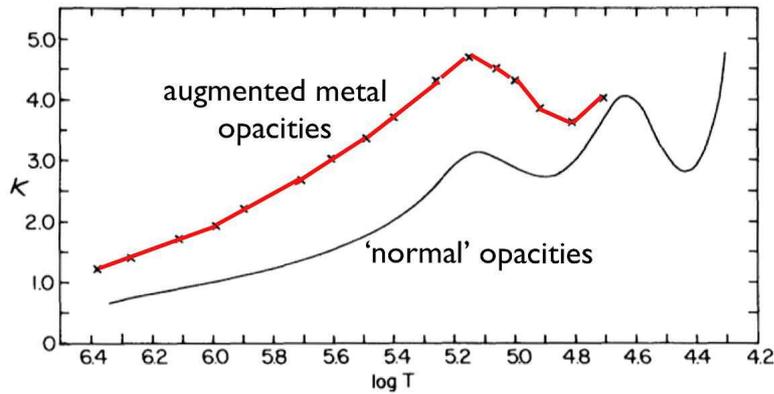}
\caption{Interior opacity in a Cepheid model from the Los Alamos compilation \citep{coxtab76}, along with the suggested enhancement needed to reconcile evolutionary and pulsation models of Cepheids, from \citet{simon82}. \label{simonopac}}
\end{figure}

In 1981 and 1982,  Norman Simon made a bold statement that these problems with Cepheid masses could be solved if the opacity caused by heavy elements was being underestimated by a factor of 2-3 in the then-current generation of opacity calculations \citep{simon81, simon82}.  He found that by making such an increase, the resulting models were able to explain the observed periods and period ratios with the same mass as required by evolutionary calculations.   This same augmentation also would provide a driving mechanism for the pulsating B stars known as $\beta$ Cepheids, another ``famous problem'' of the era.  His 1982 paper, ``{\it A Plea for Reexamining Heavy Element Opacities in Stars}'' states
\begin{verse}
It is thus quite possible that, by the single stroke of augmenting the heavy element opacities by factors of 2-3, we can bring into line with the theory of stellar structure and evolution not only the double-mode and bump Cepheids, but the $\beta$ Cepheid pulsators as well.
\end{verse}
Simon's augmented opacities from 1982 are shown in Figure \ref{simonopac}, from \citet{simon82}.  The enhancements at temperatures from 10$^{4.8}$ and above, he postulated, should result from a more careful treatment of b-f and b-b transitions in heavy elements than were implemented in the standard opacity tables available at that time.  The augmented opacities in Figure \ref{simonopac} produced models that fit the beat and bump Cepheid phenomena, and solved other problems in stellar pulsation theory.  

\begin{figure}
\includegraphics[scale=0.35]{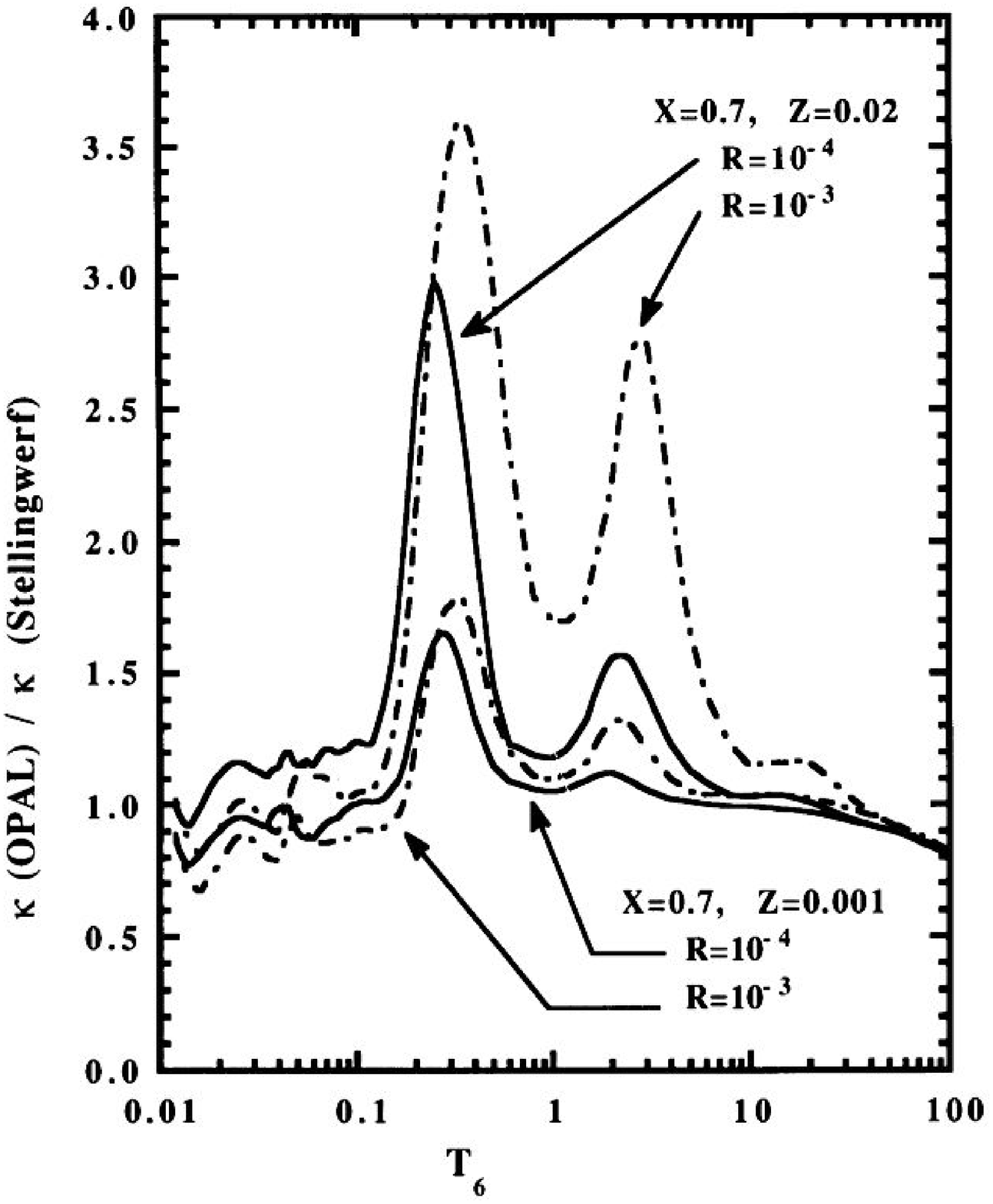}
\caption{Early OPAL opacities relevant to Cepheid interiors, compared to analytic fits by \citet{stel74, stel75} to the Los Alamos compilation \citep{coxtab76}, from \citet{iglrog91}. \label{opalceph}}
\end{figure}

This suggestion, in part, motivated Carlos Iglesias and Forrest Rogers, of Lawrence Livermore National Laboratories, to undertake a new computation of astrophysical opacities.  They eventually turned the computational machinery of their laboratory onto the problem, and found that indeed, just as predicted by Simon nearly a decade earlier that inclusion of an improved treatment of the physics of bound states in heavy elements led to an enhanced opacity.  Figure \ref{opalceph} shows that, as \citet{simon82} predicted, the opacity enhancement was a factor of 2-3 at the temperatures in that earlier work.

Using the (then) new OPAL opacities, \citet{mosetal92} revisited pulsation and evolution models of Cepheids (and $\beta$ Cepheids) to verify that the OPAL opacities did indeed solve the Cepheid mass problem.  Their results provided a dramatic closure to the problem, confirming that the newer input physics resulted in a much closer match between the masses of Cepheids from 
\begin{figure}
\includegraphics[scale=0.33]{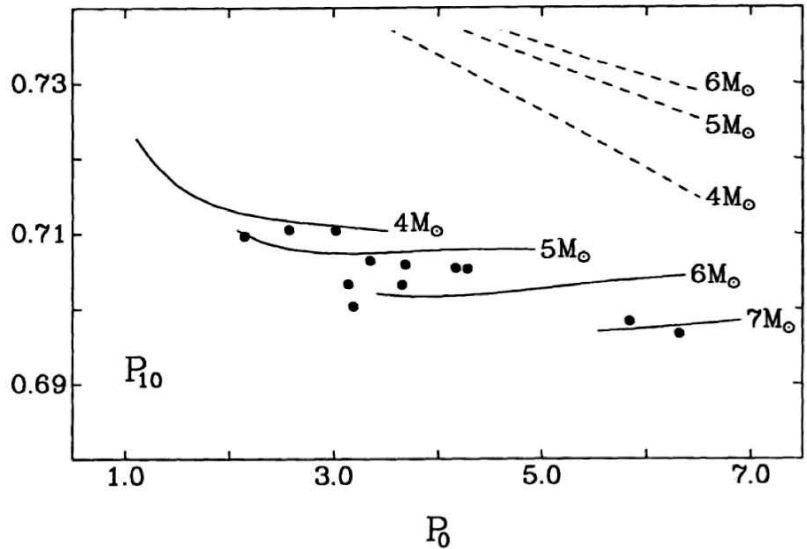}
\caption{Period ratios for beat Cepheid models compared with observations, from \citet{mosetal92}.  Dashed lines correspond to models with older opacities, solid lines are tracks for models with OPAL opacities. \label{pawelceph}}
\end{figure}
evolutionary consideration with the masses determined through pulsation period and period ratio matching.  Figure \ref{pawelceph}, for example, shows tracks for Cepheid models with the Los Alamos opacities (dashed lines) falling well above the observed beat Cepheid period ratios in the $P_{10}$-$P_{0}$ plane, while models with the OPAL opacities fit the data quite closely and at reasonable masses.  \citet{mosetal92} also show concordance for the bump Cepheids.

\section{Summary}

The previous section gave only a few illustration of the ways that stellar pulsation and asteroseismology have been able to teach us some basic physics.
Many other examples are in the current literature and form the core of active research in asteroseismology.  We haven't even mentioned the impact that helioseismology has had on the discovery and characterization of neutrino oscillations \citep{bahulr88, bahetal02} or on probes of the details of convection in the outer layers of the Sun such as in \citet{jcdetal91}.  Details of convective efficiency in white dwarfs, too, can be exposed through analysis of the shape of the pulsations \citep{mikemon07,monetal10}.  Even more fun is the possibility that measuring the evolution rate of white dwarfs could reveal the presence of axions or other exotic particles, as suggested by \citet{agnes08} and others.

These results are meaningful for our selfish purposes of improving stellar models for application to other areas of astrophysics, but also have value well outside the core astrophysics disciplines -- atomic physics, nuclear physics, and condensed matter physics in particular.   All this from a handful of coefficients in four differential equations!

\paragraph{Acknowledgements}
It was an absolute delight to have the opportunity to participate as an instructor at the Canary Islands Winter School in Astrophysics.  The IAC staff did an exemplary job in making things run smoothly, and the students who attended provided stimulating questions and discussions.  Pere Pall\'e in particular put together a great program and kept all of us instructors on task and on time.  He also provided excellent advice for all aspects of our time in La Laguna, Casa Peter most notably.  


 \end{document}